\documentclass[12pt,preprint]{aastex}

\shorttitle{\ion{N}{4}] Emitting Quasars}

\begin{document}
\title{Discovery of Two Spectroscopically Peculiar, \\ Low-Luminosity
Quasars at $z\sim 4$\altaffilmark{1}}

\author{Eilat Glikman\altaffilmark{2}, S.~G. Djorgovski\altaffilmark{2},
Daniel Stern\altaffilmark{3}, Milan Bogosavljevi\'{c}\altaffilmark{2}, \&
Ashish Mahabal\altaffilmark{2}}

\altaffiltext{1}{The data presented herein were obtained at the W.M. Keck
Observatory, which is operated as a scientific partnership among the
California Institute of Technology, the University of California and the
National Aeronautics and Space Administration. The Observatory was made
possible by the generous financial support of the W.M. Keck Foundation.}

\altaffiltext{2}{Astronomy Department, California Institute of Technology,
Pasadena, CA, 91125, email: [eilatg,george,milan,aam]@astro.caltech.edu}

\altaffiltext{3}{Jet Propulsion Laboratory, Mail Stop 169-506,
California Institute of Technology, Pasadena, CA 91109,
email: stern@thisvi.jpl.nasa.gov}

\begin{abstract}

We report the discovery of two low-luminosity quasars at $z \sim 4$,
both of which show prominent \ion{N}{4}] $\lambda$1486\AA\ emission.
This line is extremely rare  in quasar spectra at any redshift;
detecting it in two out of a sample of 23 objects (i.e., $\sim 9\%$
of the sample) is intriguing and is likely due to the low-luminosity,
high-redshift quasar sample we are studying.  This is still a poorly
explored regime, where contributions from associated, early starbursts may
be significant.  One interpretation of this line posits photoionization
by very massive young stars.  Seeing \ion{N}{4}] $\lambda$1486\AA\
emission in a high-redshift quasar may thus be understood in the context
of co-formation and early co-evolution of galaxies and their supermassive
black holes.  Alternatively, we may be seeing a phenomenon related to the
early evolution of quasar broad emission line regions.  The non-detection
(and possibly even broad absorption) of \ion{N}{5} $\lambda$1240\AA\ line
in the spectrum of one of these quasars may support that interpretation.
These two objects may signal a new faint quasar population or an early
AGN evolutionary stage at high redshifts.

\end{abstract}

\keywords{quasars: emission lines --- quasars: general --- galaxies:
evolution --- quasars: individual (DLS1053$-$0528) --- quasars: individual
(NDWFS1433+3408)}

\section{Introduction}

A growing body of theoretical models backed by emerging observational
evidence paints a picture of joint formation and evolution of
galaxies and quasars; for reviews see, e.g., the proceedings edited
by \citet{Ho04}, \citet{Djorgovski05}, and references therein.
In this picture, merger-driven build-up of massive halos trigger both
vigorous star formation and fuel the central black hole igniting a
quasar \citep{Kauffmann00,Hopkins06}.  AGN feedback is now believed
to be an essential factor in the formation and evolution of galaxies.
Understanding the AGN-starburst connection is especially interesting
at high redshifts, as we probe the epoch of the initial assembly of
massive galaxies.  Observationally, this is challenging in the case
of host galaxies of luminous Type~I quasars, as the AGN activity far
outshines any star formation related processes.  However, by observing
lower luminosity Type~I quasars or Type~II (e.g., obscured) quasars, the
blinding intensity of the central engine is reduced, allowing properties
of the host galaxy to be studied.

In this Letter, we report the discovery of two low-luminosity Type~I
quasars where we may be seeing evidence for a simultaneous starburst
with a top-heavy IMF, and AGN activity.  Both quasars show a prominent,
moderately-broad \ion{N}{4}] $\lambda 1486$\AA\ emission line, which is
rarely seen in quasar spectra at any redshift.

\section{Observations}

\begin{figure}
\epsscale{.60}
\plotone{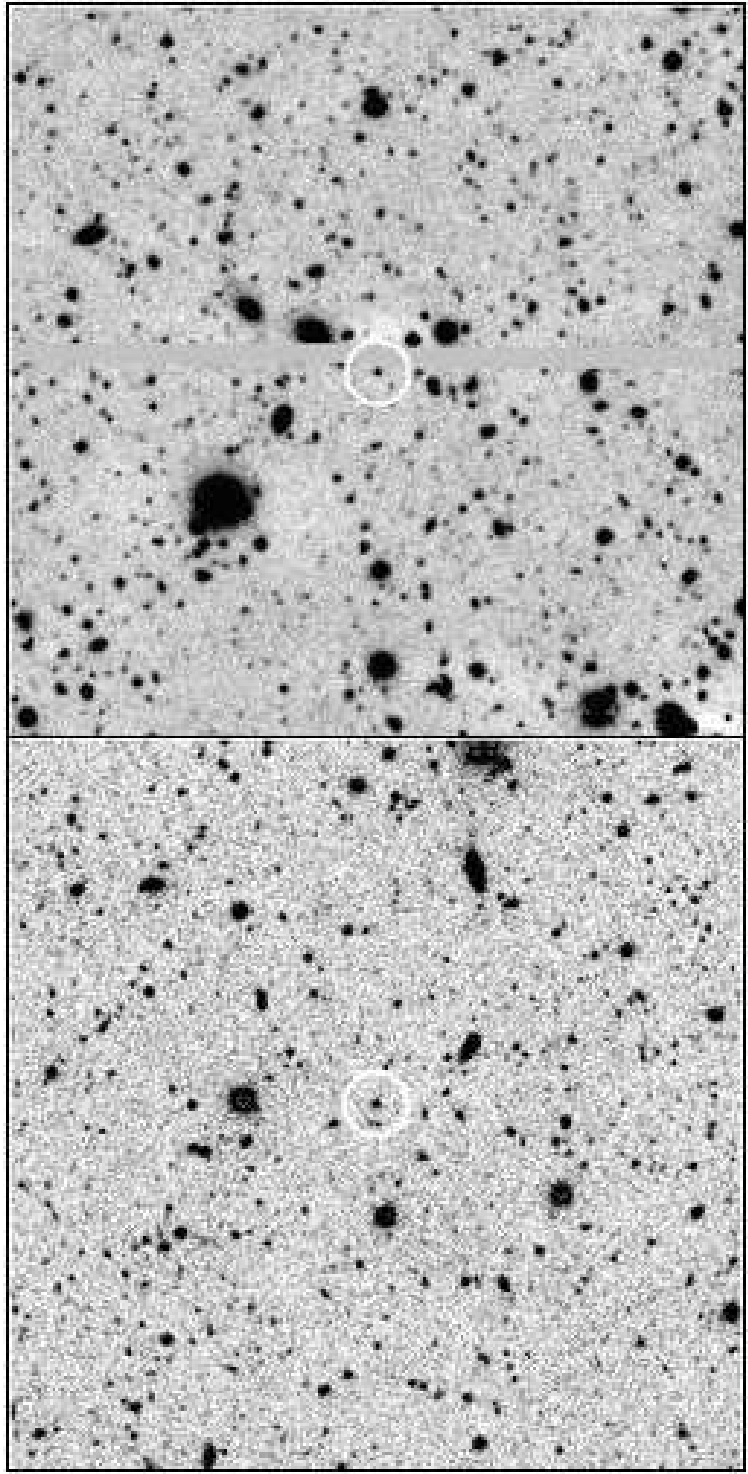}
\caption{3\arcmin $\times$ 3\arcmin\ finding charts from the $R$-band
images of the original survey data.  DLS1053$-$0528 is on the {\em top}
(the gray stipe through the center of the image masks a bleed trail from
a nearby saturated star) and NDWFS1433$+$3408 is displayed on the {\em
bottom}.  In these images, north is up and east is to the left.  The white
circles mark the locations of the quasars.}\label{fig:finding_charts}

\end{figure}

The two objects are low-luminosity quasars at $z \sim 4$, found in
a sample of 23 objects used to measure the low-luminosity end of the
quasar luminosity function (QLF) at $z \sim 4$ \citep{Bogosavlevic07}.
The parent sample ranges in redshift from $z=3.7$ to $z=5.1$.
The candidates for this sample were selected from the NOAO Deep Wide
Field Survey (NDWFS) Bo\"{o}tes field \citep{Jannuzi99}  and the Deep
Lens Survey \citep[DLS;][]{Wittman02}.  Quasar candidates were selected
based on the colors of simulated quasars in the $3.5<z<5.2$ redshift
range, incorporating a range of spectral slopes, Ly$\alpha$ equivalent
widths, and intervening neutral hydrogen absorbers, down to the limiting
magnitude of $R=24$.  Finding charts for the two \ion{N}{4}] quasars are
presented in Figure \ref{fig:finding_charts}.  Details of the survey and
candidate selection will be presented by Bogosavljevi\'{c} et al. (2007).

\begin{figure}
\epsscale{1.0}
\plotone{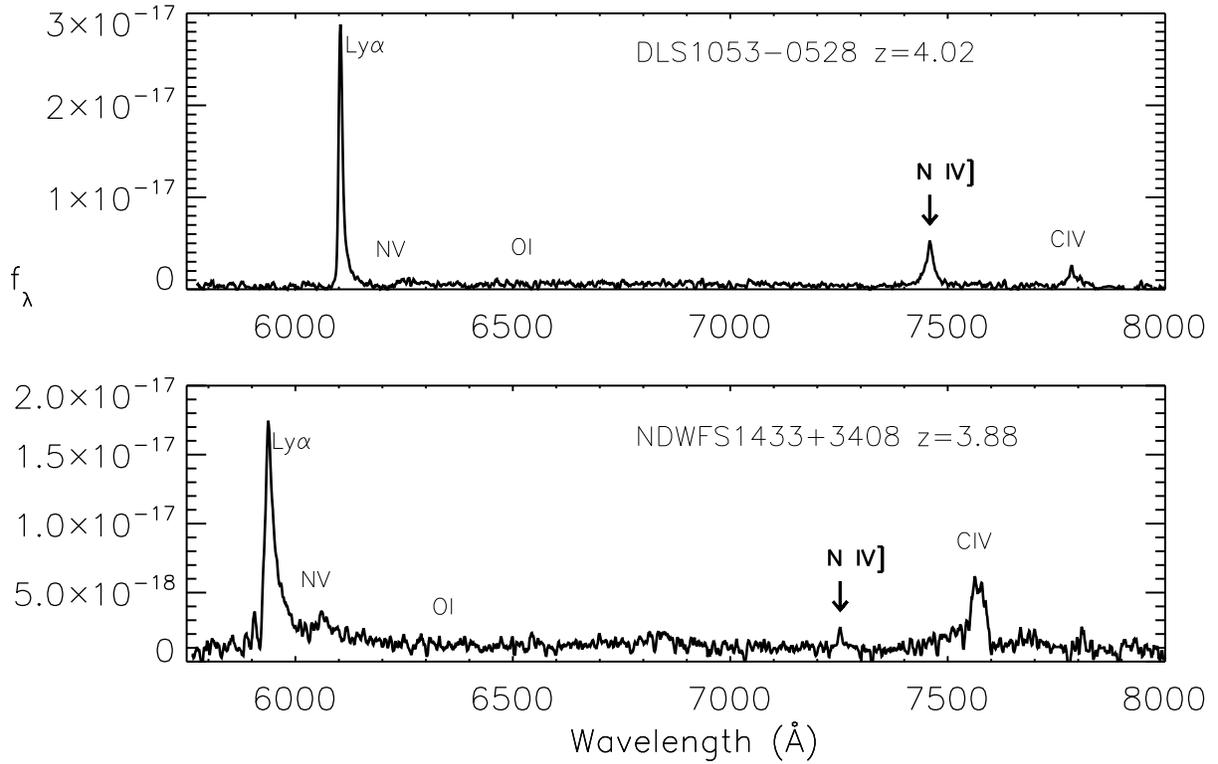}
\caption{Keck LRIS spectra of the \ion{N}{4}] $\lambda$1486\AA\
emitting quasars.  {\em Top}:  $z=4.02$ quasar found in the DLS field.
This spectrum {\em lacks} \ion{N}{5} $\lambda 1240$\AA\ in emission
but the line appears present as a broad absorption line.  {\em Bottom}:
$z=3.88$ quasar found in the NDWFS Bo\"{o}tes field.  This object shows a
weaker \ion{N}{4}] $\lambda 1486$\AA\ line and a possible weak detection
of \ion{N}{5} $\lambda 1240$\AA.}\label{fig:spectra} \end{figure}

We obtained spectroscopic followup for our candidates on UT 2006
May 20 through 22 with the Low-Resolution Imaging Spectrometer
\citep[LRIS;][]{Oke95} on the Keck I telescope. Only the red camera was
used, with the 400 lines mm$^{-1}$ grating blazed at 8500\AA.  The spectra
were reduced using BOGUS, an IRAF package developed by Stern, Bunker, \&
Stanford\footnote{\url{https://zwolfkinder.jpl.nasa.gov/$\sim$stern/homepage/bogus.html}}
for reducing slitmask data, modified slightly for our single-slit data.
DLS1053$-$0528 was discovered in a 1800 second exposure of a candidate in
the DLS F4 field and NDWFS1433$+$3408 was identified from a 900 second
exposure of a candidate in the Bo\"{o}tes field. The final spectra are
presented in Figure \ref{fig:spectra} and the source parameters are
listed in Table \ref{table:data}.

Several things are worth noting.  First is the detection of the
\ion{N}{4}] $\lambda 1486$\AA\ emission lines, which is particularly
strong in DLS1053$-$0528 (e.g., in comparison with the \ion{C}{4} $\lambda
1549$\AA\ line).  At the same time, the commonly observed \ion{N}{5}
$\lambda 1240$\AA\ emission line is entirely absent, and may even be
seen in absorption in DLS1053$-$0528 (see Figure \ref{fig:niv_fits}).
This latter observation is somewhat unusual since the \ion{C}{4}
$\lambda 1549$\AA\ line in this source does not show broad absorption.
While broad absorption lines are seen in approximately 10\% of quasars,
such quasars will show broad absorption in all permitted species.

The permitted emission lines for the two \ion{N}{4}] quasars are narrow
compared with typical quasar line-widths.  A single component Voigt fit
to Ly$\alpha$ measures a full-width at half-maximum (FWHM) of $\sim
500$ km s$^{-1}$.  Since the blue side of the Ly$\alpha$ profile is
absorbed, we forced symmetry in the line by mirroring the red side of
the line profile over the peak wavelength.  To search for a broad-line
component we fit a two-component Gaussian to the Ly$\alpha$ profile. For
DLS1053$-$0528, we measure a narrow-line FWHM of 434 km s$^{-1}$ and
a broad-line FWHM of 1727 km s$^{-1}$.  In NDWFS1433$+$3408 we measure
a narrow-line FWHM of 713 km s$^{-1}$ and a broad-line FWHM of 3015 km
s$^{-1}$.  Therefore, while these objects have significant contribution
from their narrow-line components ($50\%$ of the line flux in DLS1053$-$0528 
and $24\%$ in NDWFS1433$+$3408), their broad-line velocities of $>1000$
km s$^{-1}$ place them in the quasar regime.  Using the velocity widths
from the broad-line component of Ly$\alpha$ as a proxy for \ion{C}{4}
or H$\beta$, and the black hole mass estimators from \citet{Dietrich04}
and \citet{Vestergaard06}, we obtain $M_{BH} \sim (5 - 15) \times 10^6
~M_\sun$ for DLS1053$-$0528 and $M_{BH} \sim (32 - 77) \times 10^6
~M_\sun$ for NDWFS1433$+$3408.  Consistent with the somewhat narrow
line widths (for a broad-lined quasar) and the low-luminosity, these
inferred black hole masses are much lower than the masses inferred for
luminous quasars.

On UT 2007 January 23 we imaged DLS1053$-$0528 with the Slit-viewing
Camera (SCAM) of the NIRSPEC instrument on Keck-II.  We used the $J$-band
filter and imaged the source with a 9-point dither pattern of 120 second
exposures, for a total of 1080 seconds at the position of the quasar. The
seeing was 0\farcs7.  The data were reduced using the XDIMSUM package
in IRAF.  We calibrate our image to the 2MASS $J$-band using a 15.26
magnitude point source that is detected in the field.  The quasar is not
detected, with a 3$\sigma$ magnitude threshold of 24.9 (Vega) magnitudes.

NDWFS1433$+$3408 was observed as part of FLAMEX, a near-infrared survey
overlapping 4.1 deg$^2$ of the Bo\"{o}tes area \citep{Elston06}.  The
quasar is detected at $\sim 5\sigma$ in both the $J$ and $K_s$ images.
In addition, this quasar was observed in the IRAC Shallow Survey,
an 8.5 deg$^2$ {\em Spitzer} imaging survey of the Bo\"{o}tes field
\citep{Eisenhardt04}.  The source is faintly ($<5 \sigma$) detected
at 3.6\micron\ and 4.5\micron\ at 19.0 and 19.1 (Vega) magnitudes,
respectively.  We list the 4\arcsec\ aperture magnitude from the
FLAMEX catalog as well as the IRAC detections in Table \ref{table:data}.
The quasar is not detected in the 5.8\micron\ or 8.0\micron\ images, whose
$5\sigma$ detection thresholds are $15.9$ and $15.2$ (Vega; 3\arcsec\
aperture) magnitudes, respectively.  Note that the [3.6]$-$[4.5] colors
for this object are quite blue compared to typical mid-infrared colors
of AGN \citep[e.g.,][]{Stern05}.  The result is not unexpected:  at $z
\sim 4$, H$\alpha$ emission is shifted into the shortest wavelength IRAC
band, causing AGN at this redshift to have blue [3.6]$-$[4.5] colors.

\begin{figure}
\plotone{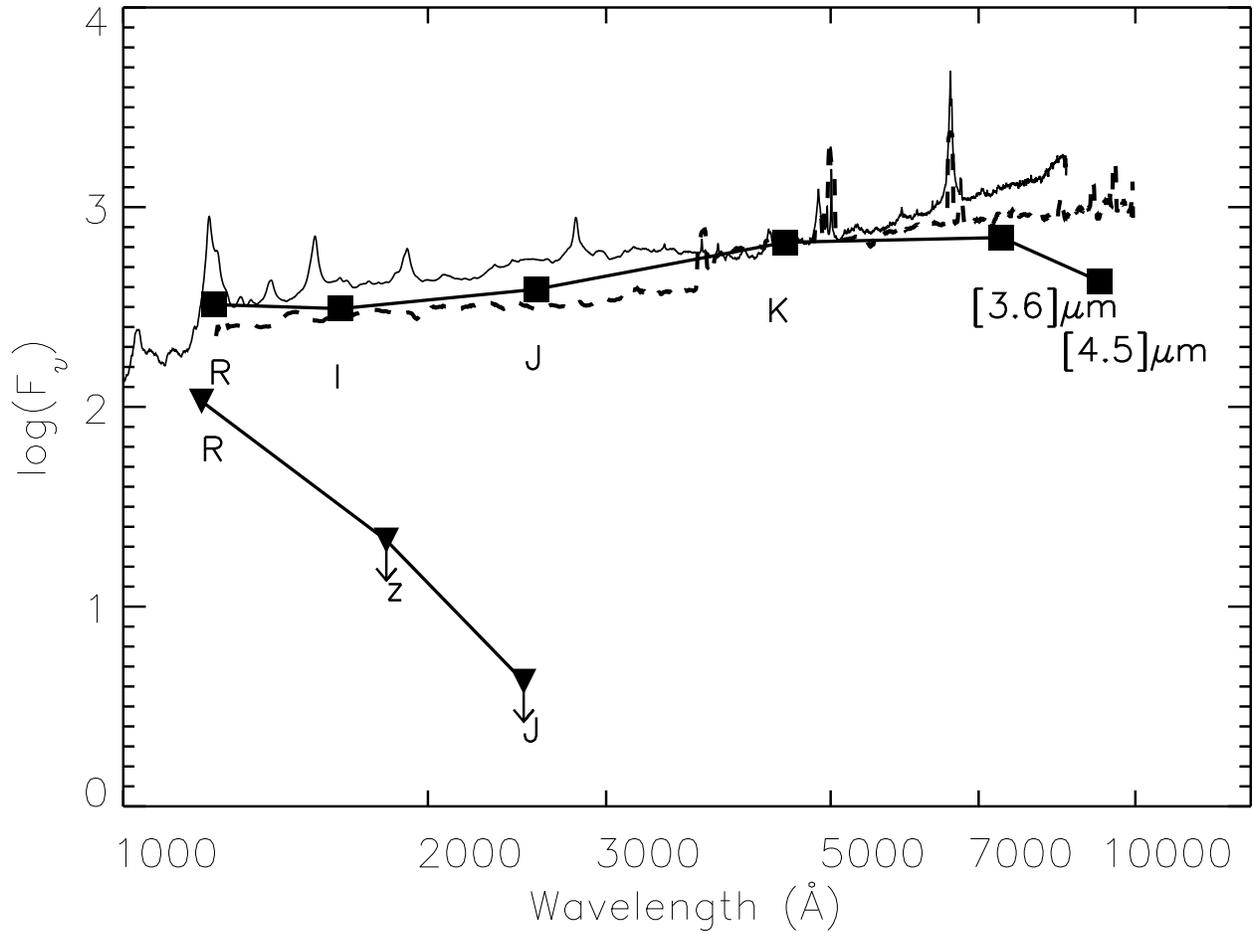}
\caption{Rest-frame spectral energy distribution (SED) for
NDWFS1433$+$3408 ({\em squares}) and DLS1053$-$0528 ({\em triangles})
compared with the SDSS quasar composite spectrum from \citet[{\em
solid line;}][]{VandenBerk01} and an unobscured starburst template with
$E(B-V)<0.1$ from \citet[{\em dashed line;}][]{Kinney96}.} \label{fig:sed}
\end{figure}

Figure \ref{fig:sed} plots the spectral energy distributions (SEDs)
of both objects, using all available photometry.  We compare this
SED with the SDSS composite spectrum \citep{VandenBerk01} as well as
the starburst template from \citet{Kinney96} with $E(B-V)<0.1$, both
shifted to $z=3.88$.  Significant deviations are seen, especially at
long wavelengths where both objects appear much  bluer than the average
quasar and starbusrt spectra.  The SED of  NDWFS1433$+$3408 seems more
consistent with the starburst spectrum, at least in the rest-frame
ultraviolet, but DLS1053$-$0528 deviates strongly from both.

\section{Discussion}

\begin{figure}
\plotone{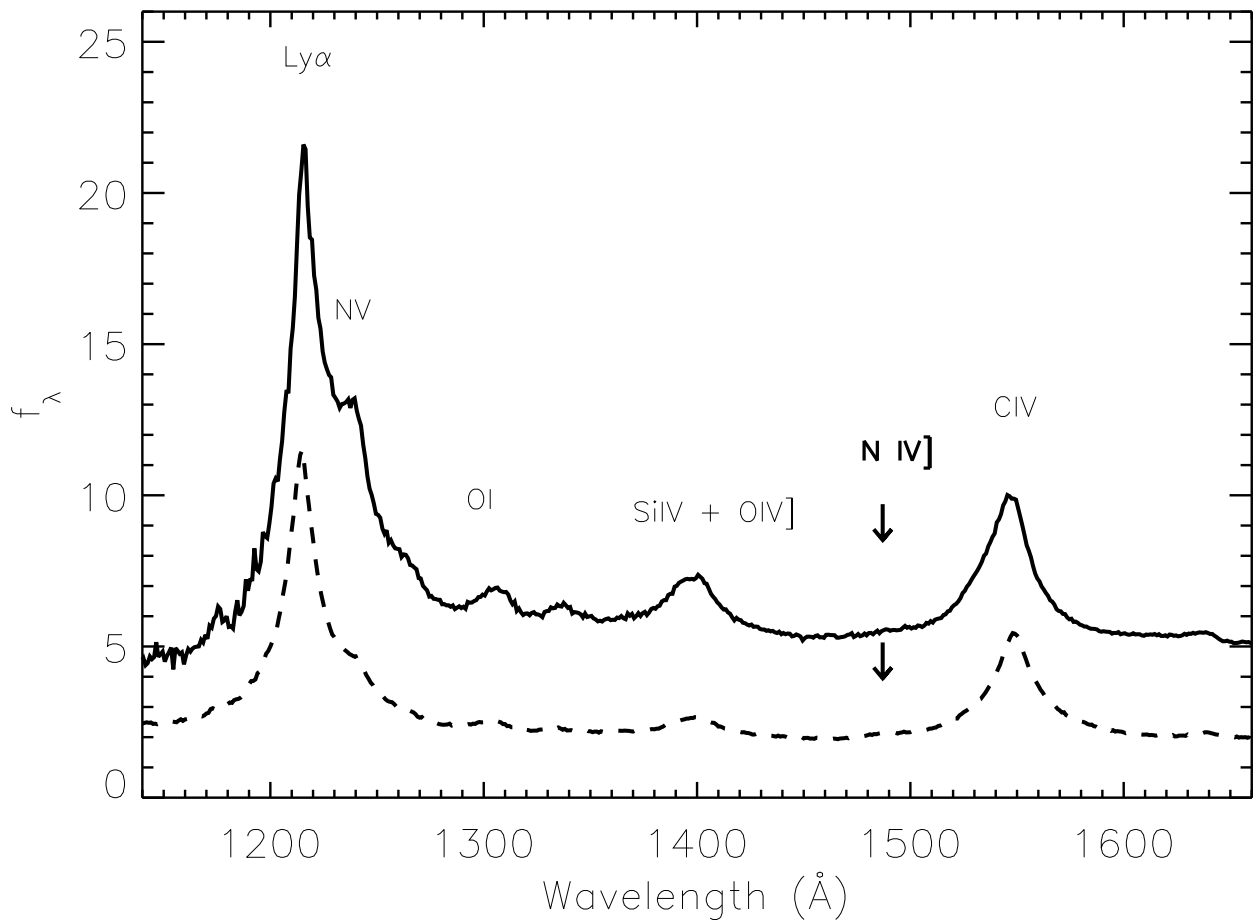}
\caption{Composite quasar spectra showing the absence of \ion{N}{4}]
$\lambda 1486$\AA.  The top spectrum (solid line) is the SDSS quasar
composite from \citet{VandenBerk01}.  The quasars that contribute to
this part of the spectrum are highly luminous high-redshift quasars.
The bottom spectrum (dashed line) is the {\em HST} UV composite spectrum
from \citet{Telfer02}. The objects contributing to this part of the
spectrum are low-redshift ($z<1$) quasars. The absence of \ion{N}{4}] in
both composite spectra suggests that it is necessary to probe deep  into
the QLF at high-reshifts to witness the interaction of star formation
with AGN activity.}\label{fig:composites} \end{figure}

The most striking feature in the spectra of these two faint quasars is the
presence of the extremely unusual \ion{N}{4}] $\lambda 1486$\AA\ emission
line.  The quasar population generally shows remarkable spectroscopic
similarity out to the highest observed redshifts, with no obvious signs of
evolution \citep[e.g.,][]{Fan06b}.  {\em None} of the published average
quasar spectral templates show any trace of the \ion{N}{4}] $\lambda
1486$\AA\ line (Figure \ref{fig:composites}).  The top spectrum (solid
line) is the SDSS quasar composite from \citet{VandenBerk01}.  The quasars
that contribute to this part of the spectrum are at redshifts comparable
to the quasars in our sample, but they sample the bright end of the quasar
luminosity function.  The bottom spectrum (dotted line) is the {\em HST}
UV composite spectrum from \citet{Telfer02}.  The objects contributing
to this part of the spectrum are low-redshift quasars ($z<1$).

\begin{figure}
\epsscale{1.0}
\plotone{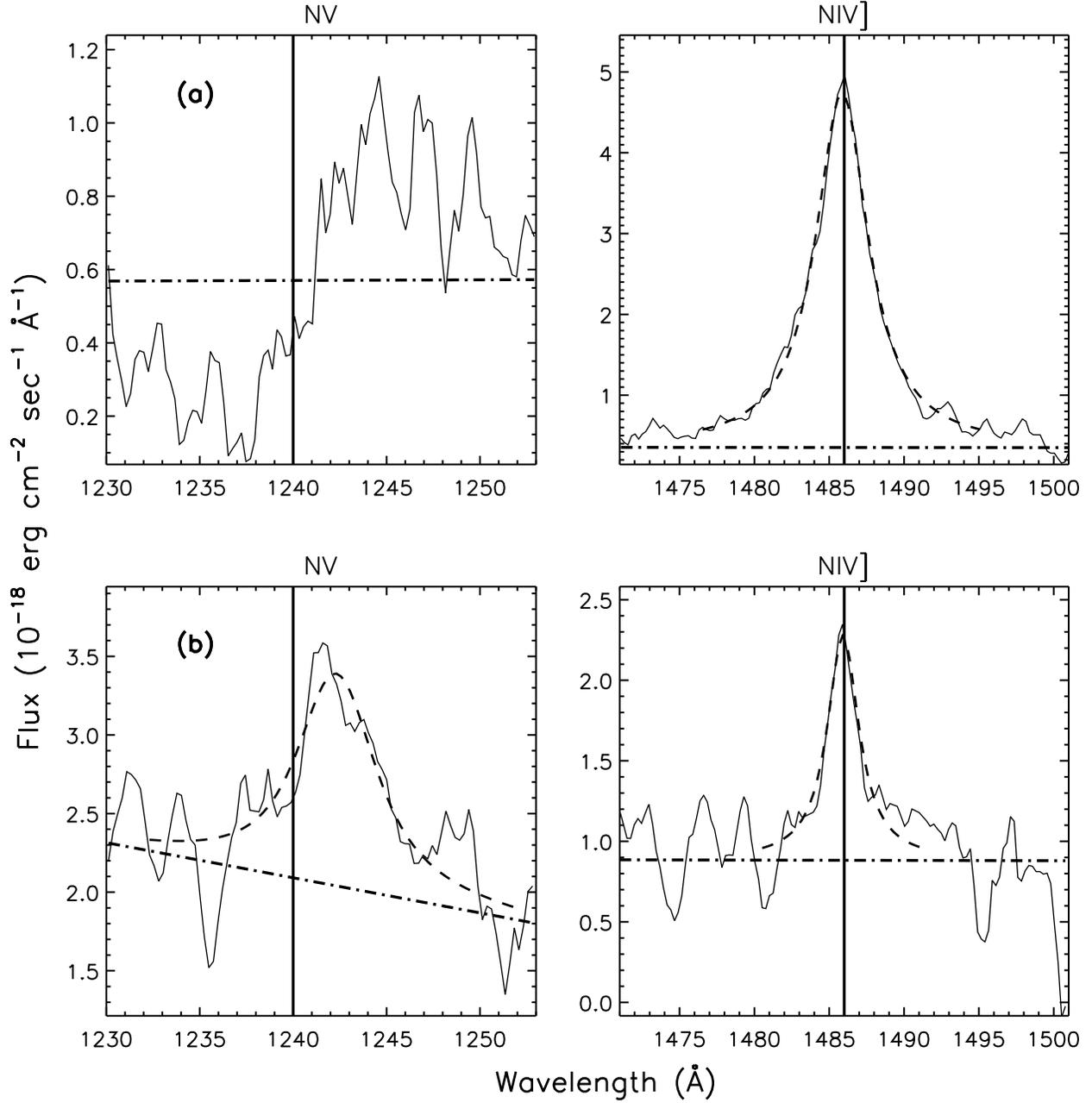}
\caption{Voigt profile fits (dotted lines, with continuum model shown
by the dot-dashed line) to Nitrogen lines in our quasars.  The line
fits to DLS1053$-$0528 are shown in the top panels and the fits to
NDWFS1433$+$3408 are shown in the bottom panels.}\label{fig:niv_fits}
\end{figure}

The only strong detection of this line in a quasar that we are aware of is
in the nitrogen quasar Q0353$-$383 \citep{Osmer80}.  This object has been
analyzed by \citet{Baldwin03} who derive an overabundance of nitrogen
by a factor as high as $5-15$ times solar in this object.  A sample
of apparently nitrogen-rich quasars was compiled by \citet{Bentz04}
from the 6650 quasars in the SDSS DR1 database with $1.6<z<4.1$
(allowing for the detection of \ion{N}{4}] $\lambda 1486$\AA\
and \ion{N}{3}] $\lambda 1750$\AA).  \citet{Bentz04} estimate that
``nitrogen-enriched'' quasars make up at most $0.2\%-0.7\%$ of quasars.
Their sample has luminosities in the range $M_i = -28.07$ to $M_i =
-24.63$.  We found two such objects in a sample of 23 quasars ($8.7\%$)
that are concentrated at $z\sim 4$ and are 0.7 magnitudes deeper than
the SDSS sample.  They found 20 quasars with equivalent widths $\geq$
2.0\AA\ in both nitrogen lines and 33 quasars with EWs $\geq$ 2.0\AA\
in only one of the nitrogen lines.  In these quasars the \ion{N}{4}]
$\lambda 1486$\AA\ line is accompanied by similarly strong \ion{N}{5}
$\lambda 1240$\AA\ and \ion{N}{3}] $\lambda 1750$\AA, the latter of which
is outside our wavelengths range.   We fit Voigt profiles to the nitrogen
lines in our spectra (as well as to Ly$\alpha$ and \ion{C}{4}  $\lambda
1549$\AA) to determine line fluxes, equivalent widths and dispersion
velocities (Table \ref{table:lines}, Figure \ref{fig:niv_fits}).  The mean
\ion{N}{4}] $\lambda 1486$\AA\ equivalent widths for the \citet{Bentz04}
sample is 3.7\AA\ ($\sigma_{EW}=$ 1.5\AA), while the equivalent widths of
\ion{N}{4}] $\lambda 1486$\AA\ for our quasars are 280.2\AA\ for the DLS
source and $24.5$\AA\ for the Bo\"{o}tes source.  This, combined with the
non-detection (and possible absorption) of \ion{N}{5} $\lambda 1240$\AA\
in the DLS source suggests that these quasars may be of a different ilk
than the ``nitrogen-enriched'' population.

In galaxies with no AGN signature, we are aware of only two other
instances of this line being seen in the high-redshift Universe. One
especially telling detection is in the spectrum of the Lynx arc,
a gravitationally lensed HII galaxy at z=3.357 \citep{Fosbury03}.
The line intensities in this object's spectrum, strong \ion{N}{4}]
$\lambda\lambda 1483,1487$\AA, \ion{O}{3}] $\lambda\lambda 1661,1666$\AA,
and \ion{C}{3}] $\lambda\lambda 1907,1909$\AA, as well as the absence of
\ion{N}{5} $\lambda 1240$, favor a hot (80,000 K) blackbody over an AGN
as the ionizing source.  The \citet{Fosbury03} modeling of the spectrum
suggests a top heavy IMF, which is especially interesting since such
an IMF is now believed to be characteristic of early, metal poor star
formation, e.g.  Population~III stars.  Alternatively, \citet{Binette03}
suggest an obscured AGN as the photoionizing source of the Lynx arc.
Their model invokes dense absorbing gas near a central AGN that filters
the powerlaw and mimicks the hot blackbody suggested by \citet{Fosbury03}
This model does require at least a weak \ion{N}{5} $\lambda 1240$\AA\
line detection.

The second known \ion{N}{4}] $\lambda 1486$\AA\ emitter is a $z=5.55$
galaxy in the GOODS survey \citep{Vanzella06,Fontanot07}.  Similar to the
Lynx arc, this object shows an extremely strong Ly$\alpha$ line as well
as \ion{N}{4}] $\lambda 1486$\AA, while \ion{N}{5} $\lambda 1240$\AA\
is absent.  No detailed analysis has been published on this source as yet.

An alternative possibility is that we are witnessing an evolutionary
effect in the quasar broad-emission-line region.  This is suggested
by the traditional broad-line shape of \ion{N}{4} $\lambda 1486$\AA\
in DLS1053$-$0528 as well the peculiar absence (or even absorption)
of \ion{N}{5} $\lambda 1240$\AA\  in its spectrum.  Detailed modeling
of such effects is beyond the scope of this paper.

The detection of such a rare emission line in two out of 23 $z\sim 4$
quasars in our sample (i.e., $\sim 9$\% of the sample) suggests that
it occurs more commonly in low luminosity quasars at high redshifts,
a regime which we are now exploring systematically for the first time.
At these luminosities we are probing deep into the QLF where the effects
of a luminous starbursts can be detected more easily; if there is an
evolutionary trend, such phenomena may be present at high redshifts,
and not in the better studied quasar samples at $z \sim 0 - 2$.  As the
sample sizes of comparably faint AGN at these redshifts increase, we will
be able to determine whether this is indeed a systematic evolutionary
effect related to the early stages of co-formation and co-evolution of
galaxies and AGN.  \\ \\

We thank the referee for helpful feedback.  We are grateful to the staff
of W. M. Keck observatory for their assistance during our observing runs.
This work was supported in part by the NSF grant AST-0407448, and by
the Ajax foundation.  The work of DS was carried out at Jet Propulsion
Laboratory, California Institute of Technology, under a contract with
NASA.

\clearpage

\begin{deluxetable}{lcccccccc}
\tabletypesize{\footnotesize}
\tablecolumns{7}
\tablewidth{0pt}
\tablehead{
\colhead{} & \colhead{R.A.} & \colhead{Dec.} & \colhead{} & 
\colhead{$R$} & \colhead{$J$}  & \colhead{$K_s$} &
\colhead{[3.6]} & \colhead{[4.5]} \\
\colhead{Name} & \colhead{(J2000)} & \colhead{(J2000)} & \colhead{$z$}  & 
\colhead{(mag)\tablenotemark{a}} & 
\colhead{(mag)} & \colhead{(mag)} & \colhead{(mag)} & \colhead{(mag)}  
}
\tablecaption{\ion{N}{4}] Emitting Quasars \label{table:data}}
\startdata
DLS1053$-$0528   & 10 53 46.1 & $-$05 28 59 & 4.02 & 23.83 & $>$26.4 &\nodata&\nodata&\nodata \\
NDWFS1433$+$3408 & 14 33 24.5 & $+$34 08 41 & 3.88 & 22.62 &   21.49 & 19.94 & 19.0 & 19.1 \\ 
\enddata
\tablenotetext{a}{AB magnitudes.  All other magnitudes are Vega-based magnitudes.}
\end{deluxetable}

\clearpage

\begin{deluxetable}{lccc}
\tablecolumns{4}
\tablewidth{0pt}
\tablehead{
 \colhead{} & \colhead{Line Flux}                   & \colhead{rest-frame EW}    & \colhead{FWHM} \\
\colhead{Line} & \colhead{(10$^{-16}$ erg cm$^{-2}$ s$^{-1}$)} & \colhead{(\AA)} & \colhead{(km s$^{-1}$)} 
}
\tablecaption{Quasar Line Properties \label{table:lines}}
\startdata
\sidehead{DLS1053$-$0528:}
 Ly$\alpha$  & $2.92\pm0.08$   & 762.1   & 1726.8\tablenotemark{a} \\
 \ion{N}{5}  & \nodata                      & \nodata & \nodata \\
 \ion{N}{4}] & $1.048\pm0.008$ & 280.2   &  807.5  \\
 \ion{C}{4}  & $0.55\pm0.02$     &  91.1   & 1101.5 \\ 
\sidehead{NDWFS1433$+$3408:}
 Ly$\alpha$ & $3.8\pm0.3$     & 244.6   & 3015.6\tablenotemark{a} \\
\ion{N}{5}  & $0.25\pm0.18$    &  11.6   & 1086.9 \\
\ion{N}{4}] & $0.21\pm0.07$    &  24.5   &  524.9 \\
\ion{C}{4}  & $1.3\pm0.3$     &  91.1   & 1334.0 \\ 
\enddata
\tablenotetext{a}{These line widths are from the broad component of a two-component Gaussian fit to the Ly$\alpha$ line.  See \S2 for details on the fitting procedure.}
\end{deluxetable}

\clearpage

\clearpage

\clearpage

\clearpage

\clearpage

\end{document}